\documentclass{PoS}

\usepackage{graphicx}
\usepackage{dcolumn}
\usepackage{bm}
\usepackage{amssymb}
\usepackage{amsmath}
\usepackage{epsfig}

\newcommand\figdir{./}

\newcommand\Tr{{\rm Tr\,}}

\newcommand\beq{\begin{eqnarray}}
\newcommand\eeq{\end{eqnarray}}

\newcommand\Eq[1]{Eq. (\ref{eq:#1})}

\newcommand\Fig[1]{Fig. (\ref{fig:#1})}
\newcommand\Tab[1]{Table (\ref{tab:#1})}

\newcommand{\calN}{{\cal N}}
\newcommand{\calO}{{\cal O}}

\newcommand{\Dslash}{\not \hspace{-0.06cm} D}

\makeatletter

\newcommand{\Rmnum}[1]{\expandafter\@slowromancap\romannumeral #1@}
\makeatother

\title{$\calN = 1$ super Yang-Mills on the lattice}

\ShortTitle{$\calN = 1$ super Yang-Mills on the lattice}

\author{%
\speaker{Michael G. Endres}\\
Physics Department, Columbia University, New York, NY 10027, USA \\
E-mail: \email{mge2112@columbia.edu}
}

\abstract{%
We present results from a numerical study of $\calN=1$ supersymmetric Yang-Mills theory using domain wall fermions.
A set of dynamical simulations were performed for the gauge group $SU(2)$ using the Wilson gauge action on $8^3\times8$ and $16^3\times32$ lattices.
We considered a range of gluino masses (i.e., fifth dimension extents $L_s=16 - 28$ and input gluino mass values $m_f=0.01-0.04$) in order to perform chiral limit extrapolations of physical quantities.
In these proceedings, we summarize our findings from a study of the Dirac spectrum and present new results for the topological charge on $\beta=2.3$, $2.35\bar3$ and $2.4$ ensembles.
}

\FullConference{%
The XXVII International Symposium on Lattice Field Theory - LAT2009 \\
July 26-31 2009 \\
Peking University, Beijing, China
}

\begin{document}

\section{Introduction}

Supersymmetry (SUSY) is an important component of numerous proposals for beyond the standard model physics.
Theories that possess SUSY exhibit a variety of fascinating properties which are of theoretical interest in their own right.
Over the past two decades, a substantial effort has been devoted to the theoretical challenge of constructing nonperturbative descriptions of SUSY theories, and this has lead to enormous progress in the field of lattice SUSY.
More recently, as a result of algorithmic and hardware advances, there has been a concerted effort devoted toward performing numerical simulations of such theories.
Among the SUSY theories of interest--and one that is the focus of these proceedings--is $\calN=1$ supersymmetric Yang-Mills (SYM) theory.
This theory is the only four-dimensional supersymmetric gauge theory which can be simulated on the lattice without numerical fine-tuning of operators.

$\calN=1$ SYM consists of a gauge boson and a Majorana fermion (i.e., gluino), each of which transforms as an adjoint under the gauge group.
The theory is confining, and possesses a variety of interesting phenomena which are accessible to nonperturbative study via lattice simulations.
Of particular interest is whether or not a gluino condensate forms, which would signal the breakdown of a discrete chiral symmetry from $Z_{2N} \to Z_2$.
The presence of discrete chiral symmetry breaking implies the formation of domain walls, for which a known, and nontrivial relationship between the domain wall tension and condensate may be tested.
At present, very little is also known about the low-lying spectrum of the theory, which is believed to consist of glue-glue, glue-gluino and gluino-gluino composite states \cite{Farrar:1997fn,Veneziano:1982ah}.
Finally, lattice simulations can provide quantitative insights into how a soft SUSY breaking gluino mass affects such quantities as the string tension and spectrum.

With conventional lattice discretizations of $\calN=1$ SYM, the only relevant SUSY violating radiative correction which is allowed by gauge invariance and the hypercubic symmetry of the lattice is a gluino mass term.
However, because the continuum theory possesses a discrete $Z_{2N}$ chiral symmetry (which is an unbroken subgroup of the anomalous U(1) axial symmetry), the gluino mass term will be protected from radiative corrections, provided the chiral symmetry is realized on the lattice.
For this reason, domain wall fermions (DWFs) are an ideal fermion discretization--and have the added advantage that, for $\calN=1$ SYM, they are free of sign problems \cite{Kaplan:1999jn}.
With the use of DWFs, supersymmetry emerges in the continuum and chiral limits in a controlled and theoretically understood fashion without fine-tuning.

We have recently performed dynamical simulations of $\calN=1$ SYM using DWFs \cite{Endres:2009yp, Endres:2008tz}\footnote{A similar study of $\calN=1$ SYM using DWFs was recently reported in \cite{Giedt:2008xm}.}.
Our study may be, to an extent, viewed as a continuation of the early exploratory simulations of \cite{Fleming:2000fa}.
In our work, we performed measurements of basic quantities and established important benchmarks which are crucial for future studies of the theory.
These include: 1) a study of the residual chiral symmetry breaking (i.e., the residual mass $m_{res}$) and the systematic errors associated with the chiral extrapolation of the gluino condensate, 2) establishing the lattice scale and 3) performing a detailed investigation of the Dirac spectrum.

Here, we present some of our results from a study of the Dirac spectrum, as well as new results for the topological charge.
The primary motivation for studying the eigenvalues of the Hermitian DWF Dirac operator is several fold.
First, we would like to verify that the DWF formulation is working properly.
This entails verifying that the low energy modes are indeed bound to the fifth dimension boundaries, and that those modes properly describe the desired chiral physics, reflected by the nature of the  matrix elements of the physical chirality operator.
Second, the eigenvalues provide a qualitative measure of the proximity to the continuum limit; in this limit the positive and negative eigenvalues of $\gamma_5 \Dslash$ should be paired, however, this pairing is broken by the lattice spacing.
Finally, the eigenvalues provide an independent method for determining the residual mass and gluino condensate, where the latter is achieved by exploiting the Banks-Casher relation \cite{Banks:1979yr}.

By measuring the topological charge, we are able to monitor topological ergodicity in our simulations.
Furthermore, on a configuration by configuration basis we may also check whether or not the near-zero modes of the Hermitian DWF Dirac operator are consistent with the gluonic definition of the topological charge.

\section{Ensembles}

We have performed a series of numerical simulations of $\calN=1$ SYM for the gauge group $SU(2)$.
Simulations were performed using DWFs \cite{Kaplan:1992bt,Furman:1994ky} and Wilson gauge action on $8^3\times8$ and $16^3\times32$ lattices.
Our current simulation parameters include fifth dimension extents ranging from $L_s = 16 - 28$, and input gluino mass values ranging from $m_f = 0.01 - 0.04$.
The majority of our simulations were performed at a single lattice spacing, which corresponds to $\beta=2.3$.
However, several ensembles were generated at finer lattice spacings, which correspond to $\beta=2.35\bar3$ and $\beta=2.4$.
The ensembles considered in these proceedings are a subset of those discussed in \cite{Endres:2009yp, Endres:2008tz} and are listed in \Tab{ensembles}.

\begin{table}[t]
\caption{%
Simulation parameters, gluino condensate $\langle\bar q q \rangle$, average plaquette $\langle \bar P \rangle$, residual mass $m_{res}$ and Sommer scale $r_0$ for a subset of ensembles found in \cite{Endres:2009yp,Endres:2008tz}.
Roman numerals serve as ensemble identifiers.
The Sommer scales quoted for Ensembles \Rmnum{2}-\Rmnum{4} are estimates obtained from $L_s=16$, $20$ and $24$ ensembles.
}
\centering
\begin{tabular}{c c c c c|c c c c c}
\hline\hline
          & $V\times T$    & $\beta$     & $L_s$ & $m_f$ & $\langle\bar q q \rangle$ & $\langle \bar P \rangle$  & $m_{res}$   & $r_0$  \\
\hline
\Rmnum{1} & $8^3 \times8 $ & 2.3         & 24    & 0.02  & 0.006806(34)              & 0.73202(32)               & $-$         & $-$    \\
\Rmnum{2} & $16^3\times32$ & 2.3         & 28    & 0.02  & 0.0063346(33)             & 0.731688(32)              & 0.14834(13) & 3.3(1) \\
\Rmnum{3} & $16^3\times32$ & $2.35\bar3$ & 28    & 0.02  & 0.0057106(61)             & 0.743853(38)              & 0.10269(18) & 4.3(1) \\
\Rmnum{4} & $16^3\times32$ & 2.4         & 28    & 0.02  & 0.0049179(82)             & 0.752951(27)              & 0.06513(17) & 5.3(1) \\
\hline
\hline
\end{tabular}
\label{tab:ensembles}
\end{table}

In order to relate the lattice parameters to the physical system, it is important to recall that the lightest states in $\calN=1$ SYM are the QCD analogs of glueballs, the $\eta^\prime$ and it's corresponding fermionic superpartner (for which there is no QCD analog).
The lightest states therefore have masses on the order of the inverse Sommer scale ($r_0^{-1}$), and the ratio $r_0/L$ provides a measure of the finite volume errors.
For our current ensembles, we find $r_0/L \approx 0.18 - 0.30$ for $\beta=2.3 - 2.4$, where $r_0$ in largely insensitive to $m_f$ and $L_s$.
On the other hand, the supersymmetric limit is controlled by the finite lattice spacing errors and proximity to the chiral limit.
The bare gluino mass is given by $m_g = m_f+m_{res}$, and hence the chiral regime corresponds to: $m_g r_0 <<1$.
For our ensembles, the bare gluino mass is dominated by the residual mass contribution, i.e.,  $m_{res}/m_f \sim \calO(10)$, which is strongly dependent on $\beta$ and $L_s$.
At $L_s = 28$ the gluino mass ranges from  $m_g r_0 \approx 0.35 - 0.5$ for $\beta=2.4 - 2.3$, suggesting that we are at most only moderately within the SUSY regime for our current set of ensembles.

\section{Dirac spectrum}

We measured the lowest 60 eigenvalues $\Lambda_H$ of the Hermitian DWF Dirac operator on $8^3\times8$ lattices using the method of Kalkreuter-Simma.
The details of the algorithm are described in \cite{Endres:2009yp, Blum:2001qg}.

\begin{figure}
\begin{minipage}[t]{0.486\textwidth}
\centering
\includegraphics[angle=270,width=2.9in]{\figdir 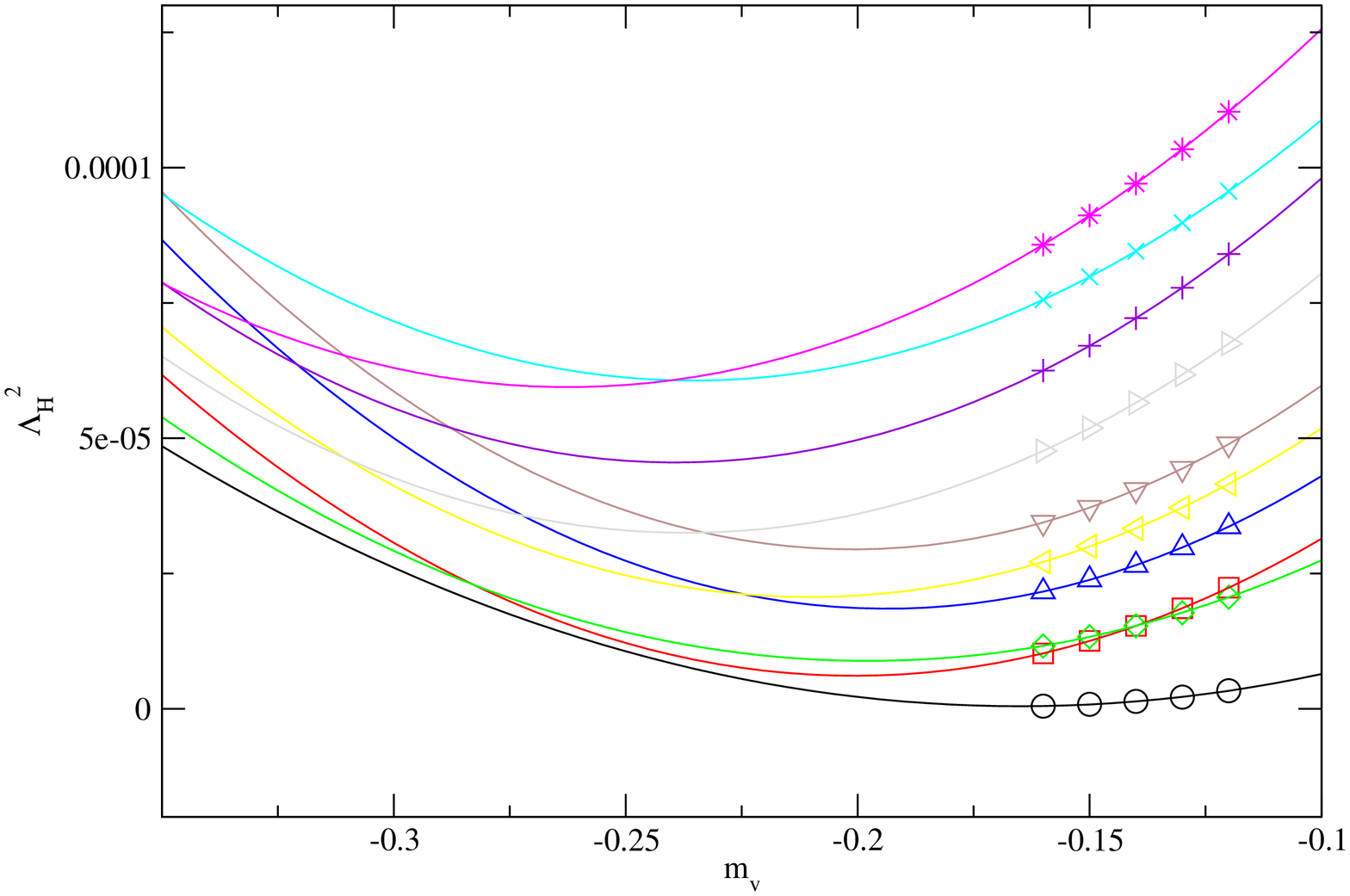}
\caption{%
Lowest 10 eigenvalues $\Lambda_H^2$ as a function of $m_v$ for a typical configuration in Ensemble \Rmnum{1}.
Solid curves represent fits to data using Eq. (3.1).
}
\label{fig:fig1}
\end{minipage}
\hspace{8pt}
\begin{minipage}[t]{0.486\textwidth}
\centering
\includegraphics[angle=270,width=2.9in]{\figdir 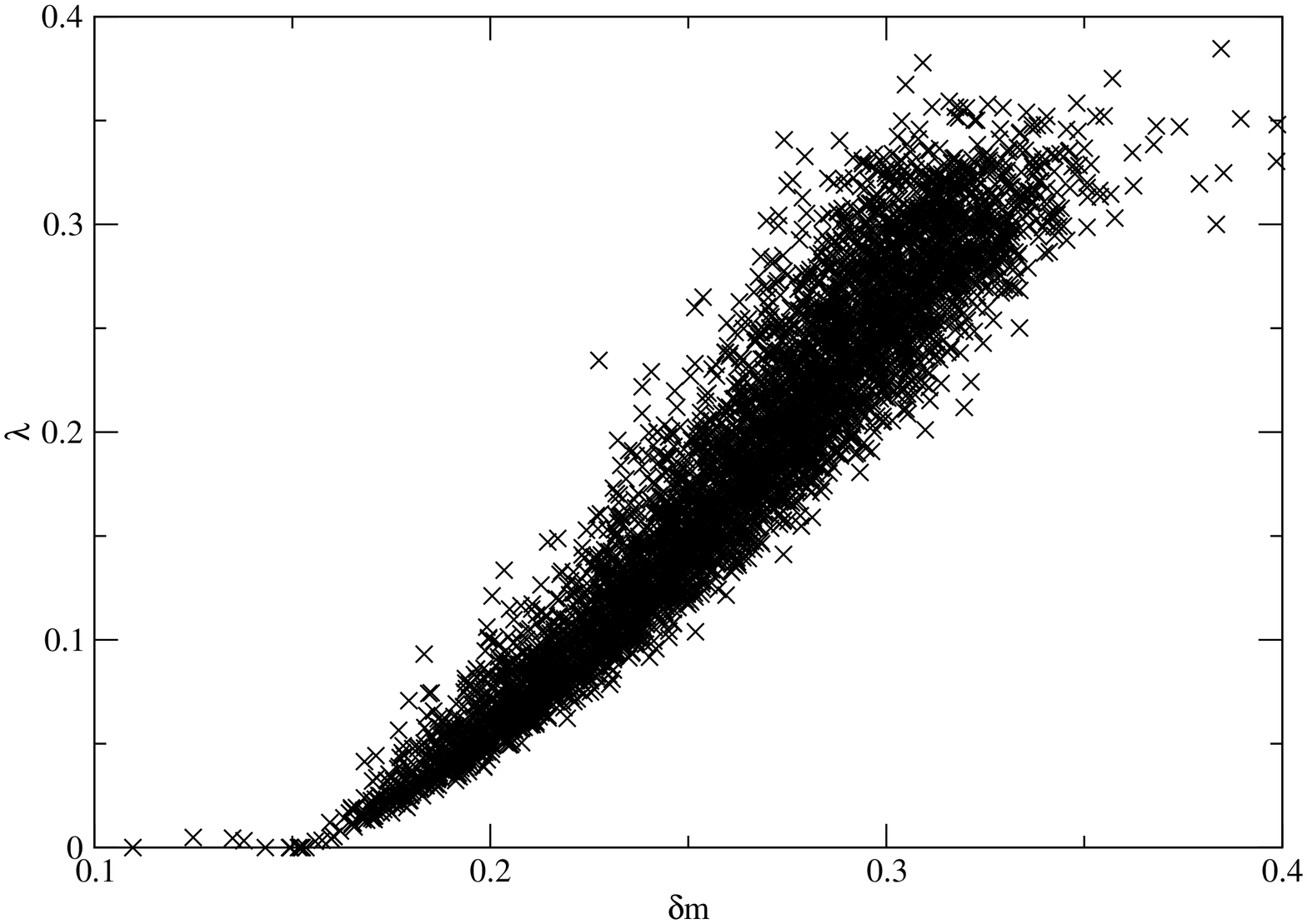}
\caption{%
Scatter plot of the fitted pairs ($\lambda$, $\delta m$) obtained from the lowest 60 eigenvalues of all configurations in Ensemble \Rmnum{1}.
}
\label{fig:fig2}
\end{minipage}
\end{figure}

The Hermitian DWF Dirac operator is given by $D_H = R_5 \gamma_5 D$, where D is the DWF Dirac operator and $R_5$ is the fifth dimension reflection operator defined in \cite{Endres:2009yp}.
We measured the eigenvalues of $D_H$ for five different values of the valence mass $m_v$, ranging from $-0.16$ to $-0.12$.
Negative values for the valence mass were chosen on the order of $m_{res}$ in order to minimize the gluino mass on each background gauge field configuration.
The valence mass dependence of the $i^{\rm th}$ eigenvalue $\Lambda_{H,i}^2$ was then fit to the reparameterized Taylor expansion \cite{Blum:2000kn}:
\beq
\Lambda_{H,i}^2(m_v) = n_{5,i}^2 \left[ \lambda^2_i + (m_v+\delta m_i)^2 \right] + \calO(m_v^3)\ .
\label{eq:fitfunc}
\eeq
\Fig{fig1} shows a plot of $\Lambda_H^2(m_v)$ and fit results for the lowest ten eigenvalues obtained from a representative configuration in Ensemble \Rmnum{1}.

In the continuum limit, the Hermitian Dirac operator $\gamma_5 (\Dslash+m_g)$ has eigenvalues $\pm\sqrt{\lambda^2+m_g^2}$, where $\pm i \lambda$ are eigenvalues of $\Dslash$ and $\lambda>0$.
The zero modes of $\Dslash$ correspond to the unpaired eigenvalues $+m$ or $-m$ of the Hermitian Dirac operator.
At finite lattice spacing, $D_H$ does not possess $\pm$ pairing of eigenvalues, and the apparent lack of pairing in \Fig{fig1} indicates that the Dirac spectrum is, at least  qualitatively, not very continuum-like.

Using the functional form of \Eq{fitfunc}, the chiral condensate may be expressed as \cite{Blum:2000kn}:
\beq
-\langle \bar q q \rangle = \frac{1}{12V} \left\langle  \sum_i  \frac{m_f+\delta m_i}{\lambda_i^2 + (m_f+\delta m_i)^2} \right\rangle \ .
\label{eq:condensate}
\eeq
The form of \Eq{condensate} allows us to identify $\lambda_i$ with the four-dimensional eigenvalues of $\Dslash$, and $\delta m_i$ as a parameter that characterize the residual chiral symmetry breaking effects on an eigenvalue by eigenvalue basis.
\Fig{fig2} shows a scatter plot of the extracted pairs $(\delta m_i,\lambda_i)$ for the lowest 60 eigenvalues of the entire ensemble.
From this plot, it is evident that there are approximately $\calO(10)$ near-zero modes for the entire ensemble, and that the majority of eigenvalues satisfy $\lambda_i << m_g$.
Note that in the regime $\lambda<<m_g$, eigenstates of $\gamma_5(\Dslash+m)$ are near-eigenstates of $\gamma_5$.
This suggests that the low-lying eigenmodes of $D_H$ are bound to the walls of the fifth dimension.
We have confirmed this behavior directly by studying the four-dimensional norms of the eigenstates of $D_H$ as a function of the fifth dimension \cite{Endres:2009yp}.

Finally, the fit parameter $\delta m_i$ characterizes the residual chiral symmetry breaking effects of finite $L_s$.
We therefore expect a relationship between the distribution of $\delta m_i$ values, $\rho(\delta m)$, and the residual mass.
Specifically, in \cite{Blum:2000kn} it was observed that $\rho(\delta m)$ was highly peaked around $m_{res}$ for quenched two flavor QCD.
Such behavior is not evident from \Fig{fig2}, indicating that the residual chiral symmetry breaking is perhaps too large to be characterized solely by $m_{res}$.

\section{Topological charge}

We measured the topological charge on $8^3\times8$ and $16^3\times32$ ensembles using both fermionic and gluonic definitions of the topological charge.
The former is given by
\beq
Q^{(f)}_{top} \propto \Tr \, D^{-1}_H \ ,
\eeq
up to a normalization factor.
In order to compare this with the gluonic definition of the topological charge, we average $Q^{(f)}_{top}$ over a smoothing window of size $n_{smooth}$.
The latter definition is determined using a two step procedure.
First, the gauge fields are smoothed via APE smearing in order to remove ultra-violet noise:
\beq
U_\mu (x) \to {\mathbb P}_{SU(2)} \left[ (1-c_{smear}) U_\mu(x) + \frac{c_{smear}}{6} \sum_{\nu\neq\pm\mu} U_\nu(x) U_\mu(x+\nu) U^\dagger_\nu(x+\mu) \right]\ ,
\label{eq:smear}
\eeq
where $U_\mu(x)$ represents a gauge link spanning sites $x$ and $x+\mu$, $c_{smear} = 0.45$ is the smearing coefficient, and the sum over $\nu$ takes both positive and negative orientations.
After each smearing step, the gauge field is projected back onto the gauge group $SU(2)$.
After a total of $n_{smear}$ applications of \Eq{smear} are performed, the topological charge is then computed using
\beq
Q^{(g)}_{top} = \frac{1}{32\pi^2}  \int d^4x\, \Tr\,  F\tilde F(x)\ ,
\eeq
where $F_{\mu\nu}$ is the five-loop improved (5li) definition of the field strength tensor built from $1\times1$, $1\times2$, $1\times3$, $2\times2$ and $3\times3$ clover-leaf terms \cite{Forcrand:1997sq}.
The use of APE smearing over the 5li cooling method of \cite{Forcrand:1997sq} was adopted by \cite{Antonio:2006px} in the context of QCD after finding that the results for the topological charge were comparable.
For convenience, we chose to use APE smearing as well, however, in the future we intend to check our analysis with cooled lattices.

\begin{figure}
\begin{minipage}[t]{0.486\textwidth}
\centering
\includegraphics[angle=270,width=2.9in]{\figdir 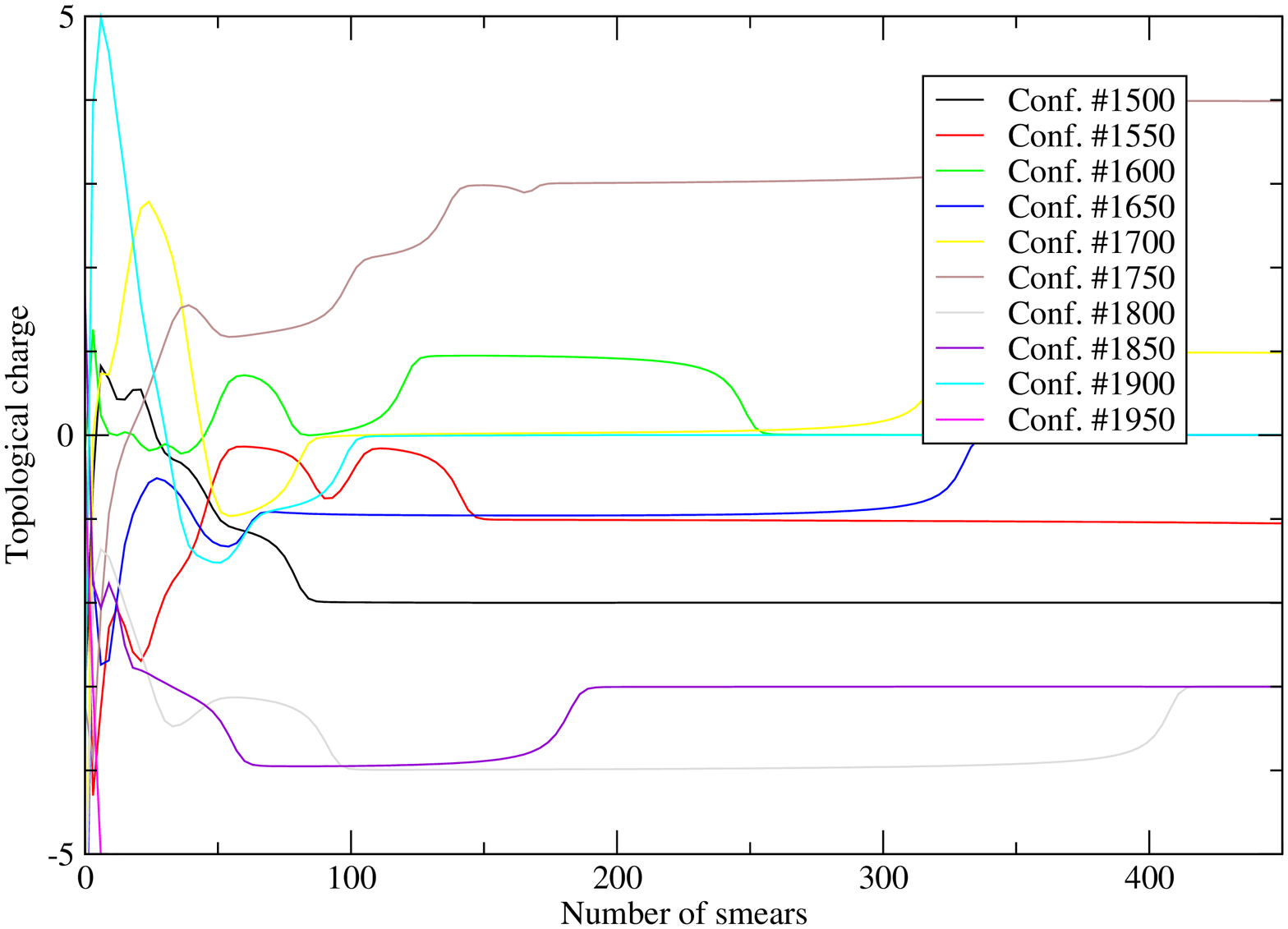}
\caption{%
Topological charge (gluonic) as a function of $n_{smear}$ for 10 different configurations in Ensemble \Rmnum{4}.
}
\label{fig:fig3}
\end{minipage}
\hspace{8pt}
\begin{minipage}[t]{0.486\textwidth}
\centering
\includegraphics[angle=270,width=2.9in]{\figdir 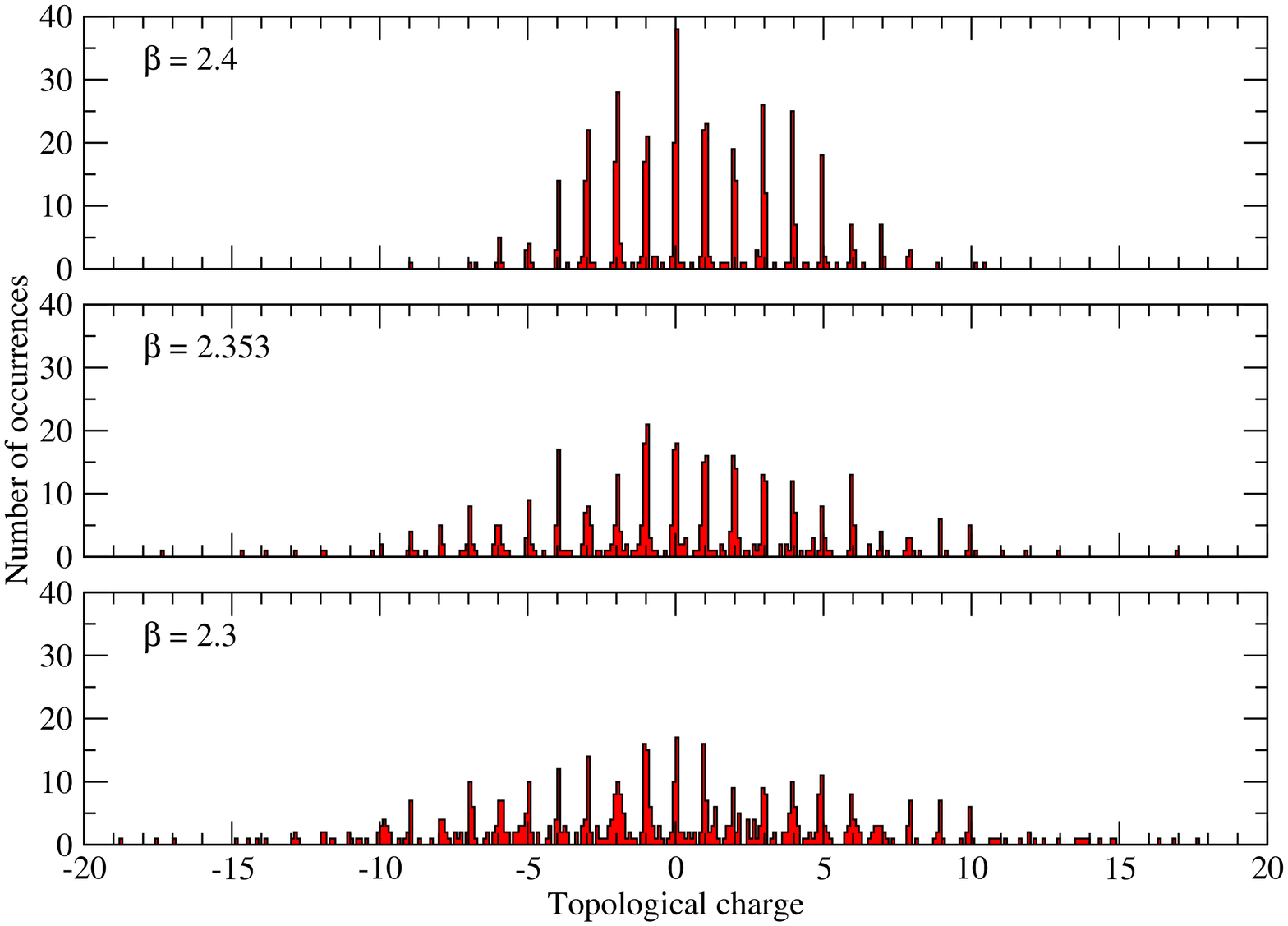}
\caption{%
Histogram of the topological charge (gluonic) evaluated at $n_{smear} = 60$ for Ensembles \Rmnum{2}, \Rmnum{3} and \Rmnum{4}.
}
\label{fig:fig4}
\end{minipage}
\end{figure}

\begin{figure}
\begin{minipage}[t]{0.486\textwidth}
\centering
\includegraphics[angle=270,width=2.9in]{\figdir 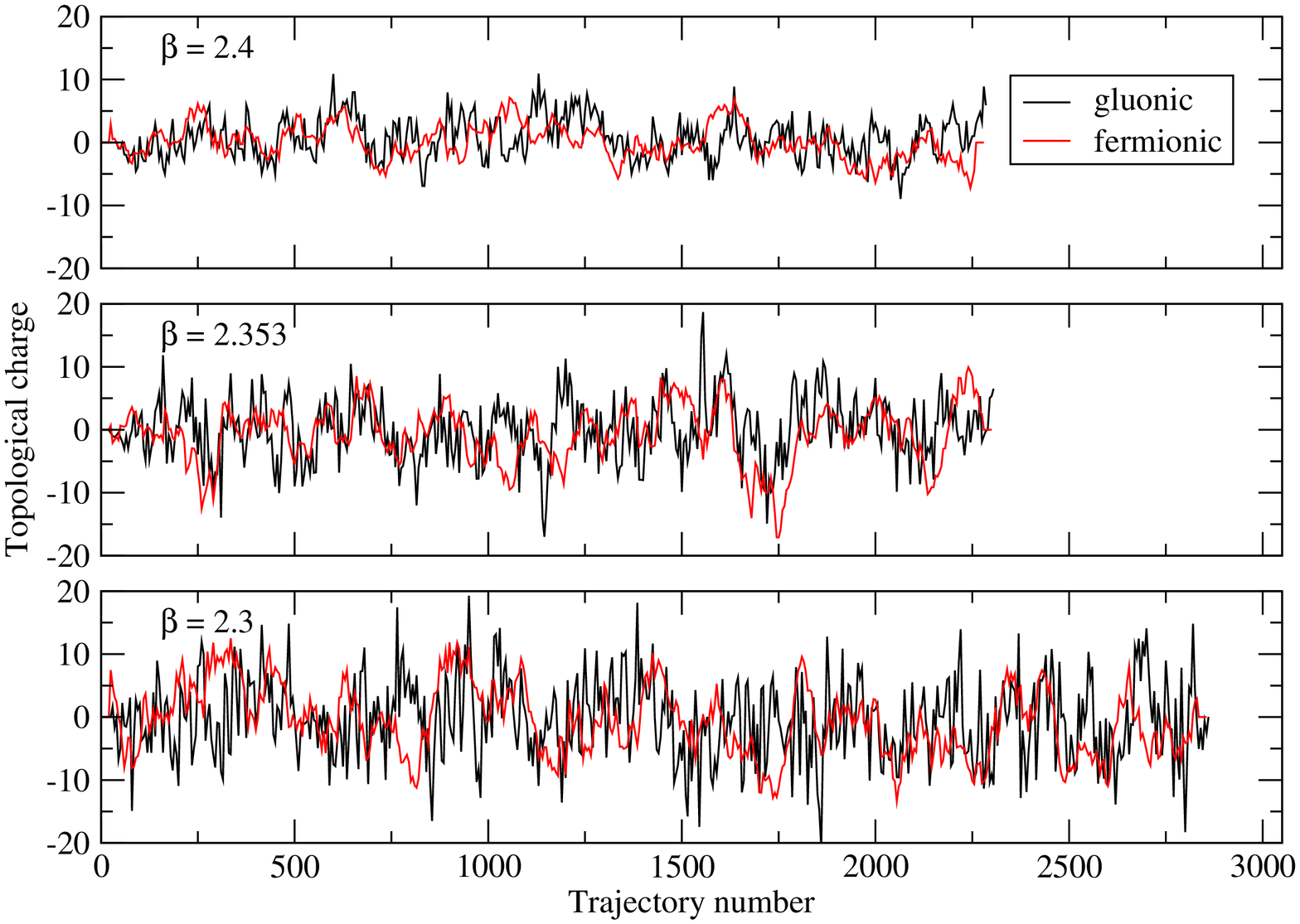}
\caption{%
Comparison of gluonic ($n_{smear}=60$) and fermionic ($n_{smooth} = 5$) definitions of the topological charge for Ensembles \Rmnum{2}, \Rmnum{3} and \Rmnum{4}.
}
\label{fig:fig5}
\end{minipage}
\hspace{8pt}
\begin{minipage}[t]{0.486\textwidth}
\centering
\includegraphics[angle=270,width=2.9in]{\figdir 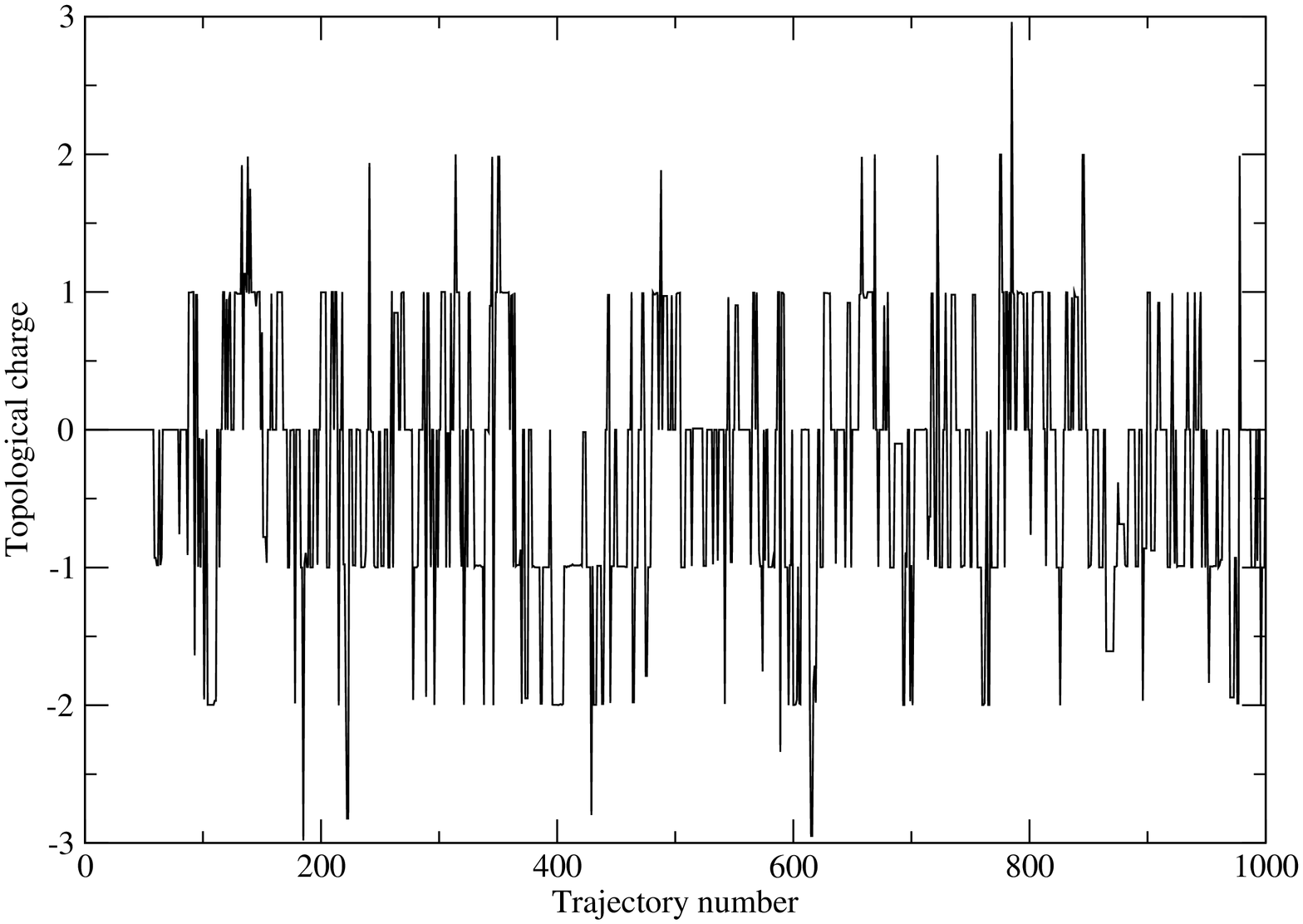}
\caption{%
Topological charge (gluonic) evaluated at $n_{smear} = 60$ as a function of trajectory number for Ensemble \Rmnum{1}.
}
\label{fig:fig6}
\end{minipage}
\end{figure}

\Fig{fig3} shows a plot of the topological charge as a function of $n_{smear}$ for some representative configurations in Ensemble \Rmnum{4}.
Beyond $n_{smear} \sim100$, the topological charge takes integer values, and infrequently fluctuates between integers.
These fluctuations are likely attributed to the disappearance of small instantons that are unstable under the APE smearing procedure.
\Fig{fig4} shows a histogram of the topological charge for three different values of the coupling (Ensembles \Rmnum{2}-\Rmnum{4}).
The Gaussian shape of the distribution indicates that the topological charge is adequately sampled in our numerical simulations.
A comparison of the fermionic and gluonic definitions of the topological charge is provided in \Fig{fig5} for all three coupling values.
The two definitions yield consistent results at the weakest coupling (Ensemble \Rmnum{4}), however at stronger couplings, the disagreement between definitions becomes increasingly evident.

On a given ensemble, we may directly compare the near-zero modes of $D_H$ with the gluonic definition of the topological charge.
\Fig{fig6} is plot of the topological charge as a function of trajectory number for the same ensemble as \Fig{fig2}, allowing for a direct test of the Atiyah-Singer index theorem.
Strictly speaking, at finite $L_s$, the DWF Dirac operator does not obey an index theorem, however, for sufficiently large $L_s$, one expects that it will do so approximately.
Based on the relatively few near-zero modes in \Fig{fig2} and the frequent sampling of non-zero topological charge sectors in \Fig{fig6}, however, there is little evidence that we are in such a regime in the case of our strongest coupling ensemble.

It would be interesting to test whether or not the index theorem is realized on our weakest coupling ensembles, since for these the comparison of fermionic and gluonic definitions of the topological charge in \Fig{fig6} exhibit greater consistency.
This consistency is a likely indication that the low-lying spectrum of $D_H$ has the desired continuum pairing of eigenvalues and that zero-modes are consistent with the topological charge.
However, because the adjoint fermion Dirac operator has two-fold degeneracy due to a charge conjugation symmetry, measuring eigenvalues on our $16^3\times32$ lattices is too costly with use of our current eigenvalue measurement code.

\section{Conclusion and outlook}

Our Dirac spectrum and topological charge results are all consistent with the existence of large residual chiral symmetry breaking attributed to finite $L_s$.
The results underscore the necessity for reducing the residual chiral symmetry breaking by going to weaker coupling, increasing $L_s$, and/or exploring improved lattice actions.
We are optimistic that with a combined approach which utilizes all three techniques, a successful study of $\calN=1$ SYM near the SUSY limit can be achieved.


\section{Acknowledgments}
M. G. E. would like to thank T. Blum, N. Christ, C. Dawson, C. Kim, R. Mawhinney and S. Takeda for useful discussions.
This research utilized resources at the New York Center for Computational Sciences at Stony Brook University/Brookhaven National Laboratory which is supported by the U.S. Department of Energy under Contract No. DE-AC02-98CH10886 and by the State of New York.
This work was supported by the U.S. Department of Energy under grant number DE-FG02-92ER40699.

\bibliography{lattice2009}
\bibliographystyle{h-physrev}

\clearpage
\pagebreak

\end{document}